\documentclass[twocolumn]{jpsj3}
\usepackage{txfonts}

\title{Magnetically Ordered State and Crystalline-Electric-Field Effects in SmBe$_{13}$}

\author{Hiroyuki Hidaka\thanks{E-mail: hidaka@phys.sci.hokudai.ac.jp}, Seigo Yamazaki, Yusei Shimizu\thanks{Present address: Institute for Materials Research, Tohoku University, Oarai, Ibaraki 311-1313, Japan}, Naoyuki Miura, Chihiro Tabata\thanks{Present address: Condensed Matter Research Center and Photon Factory, Institute of Materials Structure Science, High Energy Accelerator Research Organization, Tsukuba, Ibaraki 305-0801, Japan}, Tatsuya Yanagisawa, and Hiroshi Amitsuka}

\inst{Graduate School of Science, Hokkaido University, Sapporo, Hokkaido 060-0810, Japan} 

\abst{
The physical properties of single-crystalline SmBe$_{13}$ with a NaZn$_{13}$-type cubic structure have been studied by electrical resistivity ($\rho$), specific heat ($C$), and magnetization ($M$) measurements in magnetic fields of up to 9 T. 
The temperature ($T$) dependence of $\rho$ shows normal metallic behavior without showing the Kondo -- ln$T$ behavior, suggesting the weak hybridization effect in this system. 
Analyses of the temperature dependence of $C$ suggest that the Sm ions of this compound are trivalent and that the crystalline-electric-field (CEF) ground state is a $\Gamma_8$ quartet with a first-excited state of a $\Gamma_7$ doublet located at the energy scale of $\sim$ 90 K. 
Mean-field calculations based on the suggested CEF level scheme can reasonably well reproduce the $T$ dependence of magnetic susceptibility ($\chi$) below $\sim$ 70 K. 
These results in the paramagnetic state strongly indicate that the 4$f$ electrons are well localized with the Sm$^{3+}$ configuration. 
At low temperatures, the 4$f$ electrons undergo a magnetic order at $T_{\rm M}$ $\sim$ 8.3 K, where $\chi$($T$) shows an antiferromagnetic-like cusp anomaly. 
From the positive Curie--Weiss temperature obtained from the mean-field calculations and from a constructed magnetic phase diagram with multiple regions, we discussed the magnetic structure of SmBe$_{13}$ below $T_{\rm M}$, by comparing with other isostructural MBe$_{13}$ compounds showing helical-magnetic ordering.}

\begin{document}
\maketitle

\section{Introduction} 
The intermetallic compounds MBe$_{13}$ (M = rare earths and actinides) have attracted much interest because of the rich variety of their physical properties, such as unconventional superconductivity (SC) in UBe$_{13}$ \cite{Ott}, an intermediate-valence state in CeBe$_{13} $ \cite{Wilson}, helical-magnetic ordering in HoBe$_{13}$ \cite{Bouree}, and nuclear-antiferromagnetic (AFM) ordering in PrBe$_{13}$ \cite{Moyland}. 
Among them, UBe$_{13}$ is well known as a heavy-fermion (HF) superconductor with an extremely large electric specific-heat coefficient $\gamma$ ($\sim$ 1.1 J K$^{-2}$ mol$^{-1}$) \cite{Ott}. 
A number of studies have been conducted in order to reveal the still undetermined origin of the unconventional SC and its non-Fermi-liquid behavior in the normal phase of UBe$_{13}$ for more than thirty years \cite{Pfleiderer}. 
To obtain further insights into the novel features of UBe$_{13}$, it will be useful to reveal the common properties and differences in a series of isostructural MBe$_{13}$ compounds by studying the ground-state properties as well as magnetic correlations for each compound in more detail.

The MBe$_{13}$ compounds crystallize in a NaZn$_{13}$-type cubic structure with the space group $F$$m$$\bar{\rm 3}$$c$ (No. 226, $O_h^{\rm 6}$), where the unit cell contains M atoms in the 8$a$ site, Be$^{\rm I}$ atoms in the 8$b$ site, and Be$^{\rm II}$ atoms in the 96$i$ site \cite{Bucher, McElfresh,Takegahara}. 
It is characteristic that the unit cell consists of two \mbox{cagelike} structures: the M atom is surrounded by 24 Be$^{\rm II}$ atoms, nearly forming a snub cube, and the Be$^{\rm I}$ atom is surrounded by 12 Be$^{\rm II}$ atoms, forming an icosahedron cage. 
Recent studies of the strongly correlated electron systems with \mbox{cagelike} structures (e.g., filled skutterudites) have revealed that these systems are expected to commonly have the following characteristics: higher-order multipole degrees of freedom, strong $c$-$f$ hybridization, and low-energy phonon modes associated with the local vibration of a guest atom with a large amplitude in an oversized host cage, called rattling \cite{Aoki2, Onimaru, Yanagisawa}. 
They have provided new hot topics in strongly correlated electron physics.

The MBe$_{13}$ systems also exhibit these characteristics, which possibly originated from the \mbox{cagelike} structure; CeBe$_{13}$ is known as an intermediate-valence compound due to the strong $c$-$f$ hybridization \cite{Wilson, Lawrence}, and LaBe$_{13}$, UBe$_{13}$, and ThBe$_{13}$ have been reported to possess a low-energy phonon mode associated with the presence of a low-energy Einstein phonon mode with characteristic temperatures ($\theta_{\rm E}$) of $\sim$ 177, 151, and 157 K, respectively \cite{Hidaka, Renker, Felten}. 
In this study, we focus our attention on SmBe$_{13}$. 
Sm-based compounds have been attracting much interest because of their valence-fluctuation behavior \cite{SmS, SmB6}. 
In addition, novel phenomena have recently been found in cage-structural Sm-based compounds with cubic symmetry, such as an unusually field-insensitive HF state in SmOs$_4$Sb$_{12}$\cite{Sanada} and a magnetic-octupole ordering in SmRu$_4$P$_{12}$\cite{Yoshizawa, Aoki1}. 
It is thus intriguing to investigate the behavior of the Sm ions that are placed in the cage-structural environment.

For SmBe$_{13}$, there have been only two reports on polycrystalline samples so far \cite{Bucher, Besnus}. 
These previous works revealed the presence of a phase transition at 8.8 K, and proposed a crystalline-electric-field (CEF) level scheme for the 4$f$ electrons of Sm with a $\Gamma_7$ doublet ground state and a $\Gamma_8$ quartet first-excited state located at 12.5 \cite{Bucher} or 30 K \cite{Besnus}. 
However, the origin of the phase transition has not yet been clarified, and no attempt to grow a single crystal has been reported.
Recently, we have succeeded in growing single crystals of SmBe$_{13}$, and we performed ultrasonic measurements under high magnetic fields, which revealed the $\Gamma_8$ ground state rather than the $\Gamma_7$ ground state \cite{Mombetsu}. 
In addition, recent M$\rm \ddot{o}$ssbauer spectroscopy measurements revealed the trivalent state of the Sm ions at room temperature \cite{Tsutsui}. 
In this paper, we report the results of electrical resistivity ($\rho$), specific heat ($C$), and magnetization ($M$) measurements on single-crystalline SmBe$_{13}$.

\section{Experimental Procedure} 
Two batches of single crystals of SmBe$_{13}$, which are labeled ``$\#$1" and ``$\#$2", were grown by the Al-flux method. 
The constituent materials (Sm with 99.9$\%$ purity and Be with 99$\%$ purity) and Al with 99.99$\%$ purity were placed in an Al$_2$O$_3$ crucible at an atomic ratio of 1:13:30 and sealed in a quartz tube filled with ultrahigh-purity Ar gas of $\sim$ 150 mmHg. 
The sealed tube was kept at 1050 $\degC$ for 1 week and then cooled at a rate of 2 $\degC$/h. 
The Al flux was spun off in a centrifuge and then removed by NaOH solution. 
The obtained single crystals were annealed for 2 weeks at 700 $\degC$. 
The typical size of a grown sample is about 1 $\times$ 1 $\times$ 1 mm$^3$. 
The results of powder X-ray diffraction (XRD) at room temperature showed no impurity phase. 
The lattice parameter of SmBe$_{13}$ was obtained to be $a$ = 10.313 $\AA$, which is close to the previously reported value of $a$ = 10.304 $\AA$ \cite{Bucher}. 
In the recent single-crystal XRD measurements performed by our collaborators using synchrotron X-rays, the full width at half maximum of a rocking curve of the (200) reflection was estimated to be approximately 0.1$^\circ$, indicating the low mosaicity of the single crystals prepared in this study.

Electrical resistivity measurements using a crystal piece taken from batch $\#$1 were performed by a conventional four-probe method in the temperature range of 0.1 -- 300 K with a $^3$He/$^4$He dilution refrigerator. 
The electrical current $\boldmath j$ was applied in the [100] direction. 
Specific heat measurements were performed using a crystal piece taken from batch $\#$2 by a thermal-relaxation method in the magnetic-field range of 0 -- 9 T and in the temperature range of 2 -- 300 K with a Physical Property Measurement System (PPMS, Quantum Design, Inc.). 
DC magnetization measurements were performed using the piece taken from batch $\#$2 in magnetic fields of up to 7 T and in the temperature range from 2 to 370 K by a Magnetic Property Measurement System (MPMS, Quantum Design, Inc.). 
The weight of the sample used for the specific heat and magnetization measurements was $\sim$ 9.6 mg.

\section{Experimental Results}
\subsection{Electrical resistivity}

Figure 1 shows the temperature dependence of the electrical resistivity $\rho$($T$) of SmBe$_{13}$ for annealed and as-grown samples (batch $\#$1). 
By annealing the sample, the residual resistivity ratio (RRR) increases from 5 to 9, and the residual resistivity decreases from 7.1 to 2.7 $\mu \Omega$cm. 
In both the annealed and as-grown samples, $\rho$($T$) exhibits typical metallic behavior without any increase in $\rho$ with decreasing temperature associated with the Kondo effect, i.e., -- ln$T$ behavior, in the whole $T$ range. 
This suggests that the $c$-$f$ hybridization effect is weak in this compound.

At low temperatures, the $\rho$($T$) curve shows a kink anomaly due to a phase transition at $T_{\rm M}$, as shown in the inset of Fig. 1. 
The kink anomaly for the annealed sample is sharper and more obvious than that for the as-grown one. 
Here, the transition temperature $T_{\rm M}$ was defined as the temperature at which -- d$^2\rho$/d$T^2$ takes the maximum value. 
The obtained transition temperature $T_{\rm M}$ for the annealed sample ($\sim$ 8.7 K) is higher than that for the as-grown one ($\sim$ 6.5 K) and in good agreement with that reported previously ($\sim$ 8.8 K) \cite{Bucher}.

\begin{figure}[htb]
\begin{center}
\includegraphics[width=0.9\linewidth]{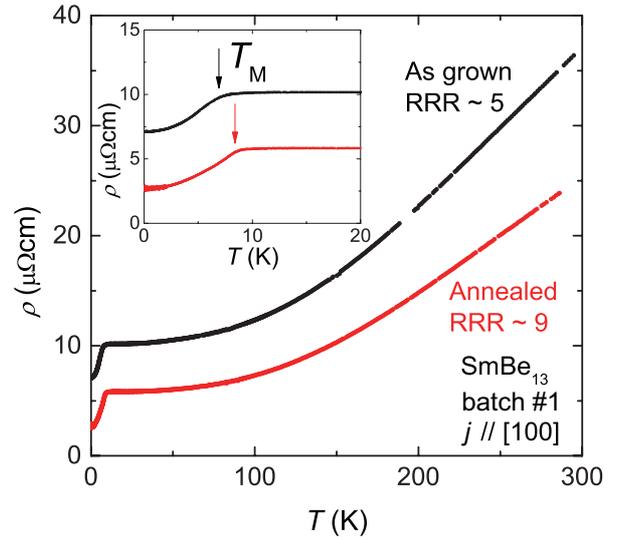}
\caption[]{\protect (Color online) Temperature dependence of electrical resistivity $\rho$($T$) of SmBe$_{13}$ for the annealed (red) and as-grown (black) samples from batch $\#$1. 
The inset shows the enlarged view below 20 K. 
The arrows indicate the transition temperature $T_{\rm M}$.} 
\end{center}
\end{figure}

\subsection{Specific heat}

Figure 2 shows the temperature dependence of the specific heat divided by the temperature $C$($T$)/$T$ for the annealed SmBe$_{13}$ (batch $\#$2) and LaBe$_{13} $\cite{Hidaka}. 
The $C$($T$)/$T$ curve of LaBe$_{13}$ obeys the Debye $T^3$ law below $\sim$ 10 K, where $\gamma$ and the Debye temperature $\theta_{\rm D}$ were estimated to be $\sim$ 9.1 mJ K$^{-2}$ mol$^{-1}$ and $\sim$ 950 K, respectively \cite{Hidaka}. 
Furthermore, the $C$($T$)/$T$ curve for LaBe$_{13}$ shows a broad hump at around 40 K, which can be well described by a simple model assuming a low-energy Einstein phonon mode with $\theta_{\rm E}$ of $\sim$ 177 K \cite{Hidaka}. 
Note that the $C$($T$)/$T$ curve of SmBe$_{13}$ also has a similar hump structure near 40 K, which should involve the CEF Schottky contribution as well as the low-energy Einstein phonon contribution, as discussed below. 
Here, the value of $C$/$T$ for SmBe$_{13}$ is slightly smaller than that for LaBe$_{13}$ above $\sim$ 80 K, which may be due to the difference in the phonon contributions.

The $C$($T$)/$T$ curve of SmBe$_{13}$ exhibits a $\lambda$-type anomaly at $T_{\rm M}$ = 8.3 K, indicating that a second-order phase transition takes place at this temperature. 
The $T_{\rm M}$ determined from the $C$ measurements is slightly lower than that obtained from the present $\rho$ measurements, which may originate from the difference in the measured sample batches. 
Below $T_{\rm M}$, a shoulder structure can be seen in $C$($T$)/$T$ (see the inset of Fig. 2), suggesting that the ordering cannot be described by a simple mean-field model with a doublet CEF ground state. 
The inset also displays $C$($T$)/$T$ near $T_{\rm M}$ in magnetic fields applied in the [100] direction. 
$T_{\rm M}$ shifts to the low-temperature side with increasing magnetic field.

\begin{figure}[htb]
\begin{center}
\includegraphics[width=0.9\linewidth]{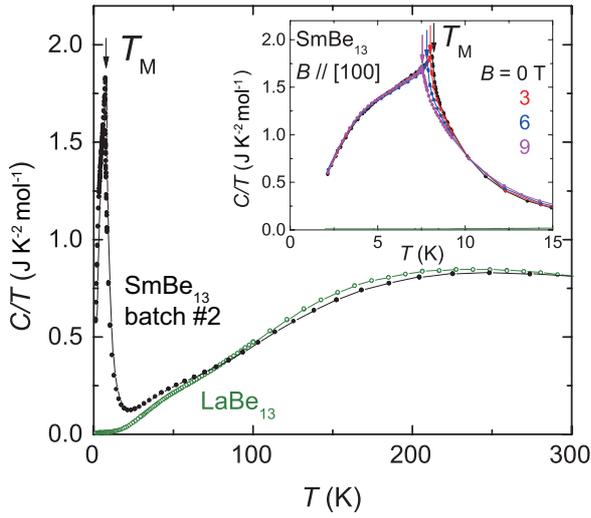}
\caption[]{\protect (Color online) Temperature dependence of $C$/$T$ for SmBe$_{13}$ (closed circles) and LaBe$_{13}$ (open circles) \cite{Hidaka} below 300 K at zero field. 
The inset shows the temperature dependence of $C$/$T$ below 15 K in various magnetic fields applied in the [100] direction. 
The arrows represent $T_{\rm M}$ determined from the peak in $C$($T$)/$T$ at each magnetic field. 
}
\end{center}
\end{figure}

Next, we estimated the contribution of 4$f$ electrons to the specific heat ($C_{\rm 4\it f}$) for SmBe$_{13}$ by subtracting $C$($T$) of LaBe$_{13}$ from that of SmBe$_{13}$, as shown in Fig. 3. 
The 4$f$-electron entropy $\Delta$$S_{\rm 4\it f}$ [$\equiv$ $S_{\rm 4\it f }$($T$) -- $S_{\rm 4\it f}$(2 K)] was obtained by integrating $C_{\rm 4\it f}$($T$)/$T$ from 2 K. 
In the cubic CEF with the $O_h$ symmetry, the $J$ = 5/2 ground multiplet of Sm$^{3+}$ splits into a $\Gamma_8$ quartet and a $\Gamma_7$ doublet. 
In SmBe$_{13}$, the $\Gamma_8$ quartet should be the CEF ground state since $\Delta$$S_{\rm 4\it f }$ reaches 0.73$R$ln4 at $T_{\rm M}$, which is significantly larger than $R$ln2. 
This is consistent with the suggested CEF ground state from the previous ultrasonic measurements \cite{Mombetsu}.
Note that the actual 4$f$-electron entropy at $T_{\rm M}$ must be even closer to $R$ln4 than $\Delta$$S_{\rm 4\it f}$($T_{\rm M}$) estimated above 2 K. 
The reduction in $\Delta$$S_{\rm 4\it f}$ from $R$ln4 at $T_{\rm M}$ is considered to be due to the occurrence of a short-range ordering above $T_{\rm M}$, since the effect of the $c$-$f$ hybridization is considered to be negligibly small, as suggested from the absence of -- ln$T$ behavior in $\rho$($T$).

At higher temperatures ($T$ $\sim$ 80 K), $\Delta$$S_{\rm 4\it f}$($T$) reaches approximately $R$ln6, which is the expected value in the $J$ = 5/2 multiplet state, suggesting that the Sm ions are trivalent. 
The trivalent state of the Sm ions for SmBe$_{13}$ has also been confirmed by M$\rm \ddot{o}$ssbauer spectroscopy at room temperature\cite{Tsutsui}. 
A broad maximum of the $C_{\rm 4\it f }$($T$) curve at around 40 K is mainly attributed to the CEF Schottky anomaly. 
Here, the phonon contribution at low temperatures in SmBe$_{13}$ is considered to be approximated by that in LaBe$_{13}$, since $\theta_{\rm E}$ estimated from our recent XRD studies does not show an obvious difference between LaBe$_{13}$ [$\theta_{\rm E}$ of 163(15) K] and SmBe$_{13}$ [$\theta_{\rm E}$ of 157(10) K]. \cite{Hidaka2} 
Assuming a $\Gamma_8$ -- $\Gamma_7$ level scheme with energy separation $\Delta_{\rm CEF}$ of 90 K, the broad maximum can be reproduced well by the calculated CEF Schottky curve in this work (the red dot-dashed line in Fig. 3). 
For the sake of comparison, we also represented the calculated $C_{\rm 4\it f }$($T$) curves based on the previously reported CEF level schemes: the $\Gamma_7$ -- $\Gamma_8$ model with $\Delta_{\rm CEF}$ = 12.5 K determined from the $C$ measurements by Bucher et al., \cite{Bucher} and the $\Gamma_7$ -- $\Gamma_8$ model with $\Delta_{\rm CEF}$ = 30 K determined from the $M$ measurements by Besnus et al. \cite{Besnus}. 
Both of them, however, cannot explain the broad maximum in $C_{\rm 4\it f }$($T$) at $\sim$ 40 K.

\begin{figure}[htb]
\begin{center}
\includegraphics[width=0.9\linewidth]{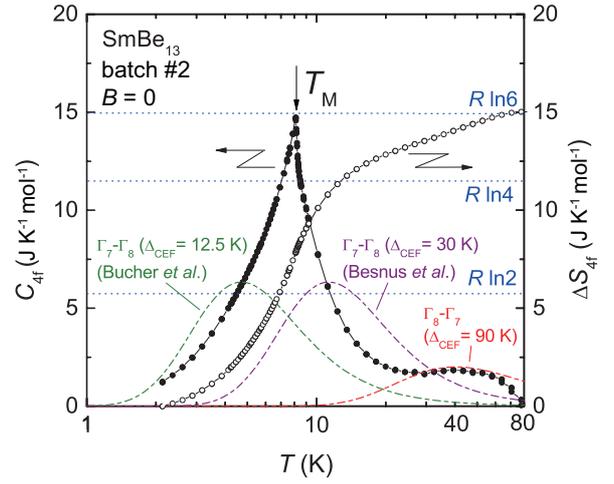}
\caption[]{\protect (Color online) Temperature dependences of $C_{\rm 4\it f }$ and $\Delta$$S_{\rm 4\it f }$ in SmBe$_{13}$ at zero magnetic field, where $\Delta$$S_{\rm 4\it f }$ $\equiv$ $S_{\rm 4\it f }$($T$) -- $S_{\rm 4\it f}$(2 K). 
The red, green, and purple dot-dashed lines represent the calculated CEF Schottky curves proposed by this study, Bucher et al.\cite{Bucher}, and Besnus et al.\cite{Besnus}, respectively. 
}
\end{center}
\end{figure}

\subsection{Magnetic susceptibility and magnetization curve}

The temperature dependence of the magnetic susceptibility $\chi$($T$) [= $M$($T$)/$B$] of SmBe$_{13}$ measured at $B$ = 0.1 T between 2 and 370 K is shown in Fig. 4. 
The measured sample was an annealed single crystal taken from batch $\#$2, and the magnetic field was applied along the [100] axis. 
$\chi$($T$) is nearly constant at high temperatures and starts increasing gradually with decreasing temperature below $\sim$ 300 K. 
Such behavior has also been reported in other Sm-based compounds, such as SmAl$_2$\cite{Wijin} and SmTi$_2$Al$_{20}$\cite{Higashinaka}, where this behavior has been interpreted to be due to the mixing of the low-lying $J$ = 7/2 excited multiplet of Sm$^{3+}$ into the $J$ = 5/2 ground multiplet. 
The increase in $\chi$($T$) for SmBe$_{13}$ becomes more pronounced below $\sim$ 100 K. 
The experimental data between 15 and 70 K can be reasonably well described in terms of the following mean-field model:
\begin{equation} 
\chi(T) = \frac{\chi_{\rm CEF}(T) }{1 - \lambda\chi_{\rm CEF}(T) } + \chi_0, 
\end{equation} 
where $\lambda$ is the mean-field constant: 
\begin{equation} 
\lambda = \frac{3 k_{\rm B} \theta_{\rm CW}}{N_{\rm A} \mu_{\rm B}^2 g_J^2 J(J + 1)}, 
\end{equation} 
$\chi_{\rm CEF}$ is the single-ion magnetic susceptibility assuming the $\Gamma_8$ -- $\Gamma_7$ CEF level scheme of the $J$ = 5/2 multiplet for Sm$^{3+}$ with $\Delta_{\rm CEF}$ = 90 K, $\chi_0$ is a constant, $k_{\rm B}$ is the Boltzmann constant, $\theta_{\rm CW}$ is the Curie--Weiss temperature, $N_{\rm A}$ is Avogadro's number, $\mu_{\rm B}$ is the Bohr magneton, $g_J$ = 2/7 is the Land\'e g factor, and $J$ = 5/2. 
From the best fit, we obtained $\chi_0$ to be 3.2 $\times$ 10$^{-4}$ emu/mol and $\theta_{\rm CW}$ to be 10.8 K. 
The good agreement between the experimental data and the mean-field calculation suggests that the magnetic-dipole moments of the 4$f$ electrons are hardly reduced below 70 K. 
The $\chi$($T$) curve deviates from the fitting curve below $\sim$ 15 K, probably due to the short-range ordering, and then it exhibits a clear cusp at $T_{\rm M}$ = 8.3 K. 
It is noteworthy that such a cusp anomaly is associated with the occurrence of AFM ordering, despite the fact that the obtained value of $\theta_{\rm CW}$ (= 10.8 K) is positive, indicating that the total effective magnetic interaction between the Sm 4$f$ moments is ferromagnetic (FM).

\begin{figure}[htb]
\begin{center}
\includegraphics[width=0.9\linewidth]{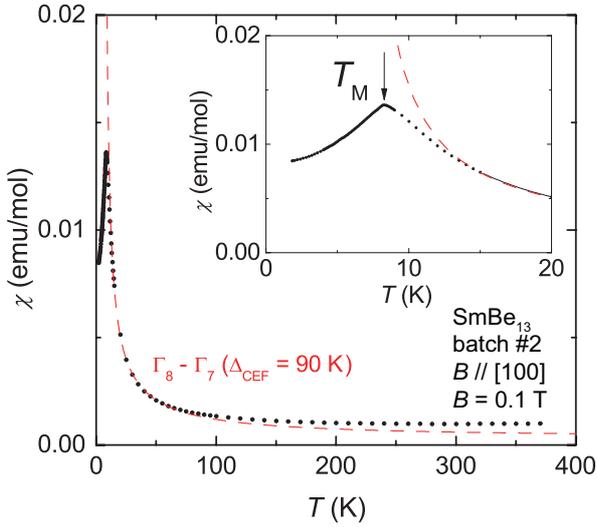}
\caption[]{\protect (Color online) Temperature dependence of magnetic susceptibility $\chi$($T$) in SmBe$_{13}$ at $B$ = 0.1 T for $B$ // [100]. The inset shows $\chi$($T$) below 20 K. The red-dashed line represents the fitting curve obtained from the mean-field calculation, as described in the text. 
}
\end{center}
\end{figure}

Figure 5 shows the temperature dependence of the magnetization $M$($T$) at various magnetic fields up to 7 T. 
The magnetic field was applied in the [100] direction. 
At each magnetic field, the results of zero-field-cooling (ZFC) and field-cooling (FC) processes are shown in Fig. 5. 
The $M$($T$) curves at low magnetic fields show a cusp anomaly at $T_{\rm M}$. 
However, the decrease in $M$ below $T_{\rm M}$ is gradually suppressed with increasing magnetic field, and then the cusp anomaly at $T_{\rm M}$ becomes a kink anomaly above 5 T. 
Here, we determined $T_{\rm M}$ from the temperature where d$^2M$/d$T^2$ shows a local minimum. 
$T_{\rm M}$ slightly decreases with increasing field, in good agreement with the results obtained from the present $C$ measurements.
In addition to the $T_{\rm M}$ anomaly, the ZFC curves above 3 T show another kink anomaly at a lower temperature than $T_{\rm M}$, defined as $T_{\rm X}$. 
$T_{\rm X}$ decreases linearly with increasing magnetic field, where $T_{\rm X}$ was also determined from the local minimum position of d$^2M$/d$T^2$. 
Moreover, there is a significant difference in $M$($T$) between the ZFC and FC curves below $\sim$ $T_{\rm X}$, suggesting the possible presence of magnetic domains.

\begin{figure}[htb]
\begin{center}
\includegraphics[width=0.9\linewidth]{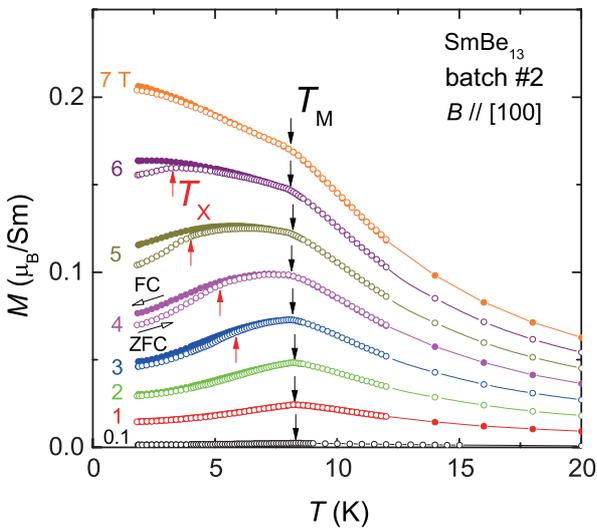}
\caption[]{\protect (Color online) Temperature dependence of magnetization for SmBe$_{13}$ measured at various magnetic fields from 0.1 to 7 T. 
The open and closed symbols represent the data obtained from the ZFC and FC processes, respectively. 
The black and red arrows indicate $T_{\rm M}$ and $T_{\rm X}$, respectively. 
}
\end{center}
\end{figure}

Figure 6 shows the magnetization curves $M$($B$) of SmBe$_{13}$, measured at various temperatures between 2 and 12 K for fields up to 7 T parallel to [100]. 
$M$($B$) at 12 K (above $T_{\rm M}$) exhibits simple paramagnetic (PM) behavior with a Brillouin curve. 
On the other hand, below $T_{\rm M}$, the $M$($B$) curves bend upward at the newly defined characteristic field $B_{\rm X}$. 
Here, we determined $B_{\rm X}$ from the intersection of two linear extrapolations from the higher- and lower-field regions. 
With decreasing temperature, the bending becomes distinct and $B_{\rm X}$ shifts to the higher-field side. 
At the lowest temperature of 2 K, $B_{\rm X}$ reaches approximately 4 T.

\begin{figure}[htb]
\begin{center}
\includegraphics[width=0.75\linewidth]{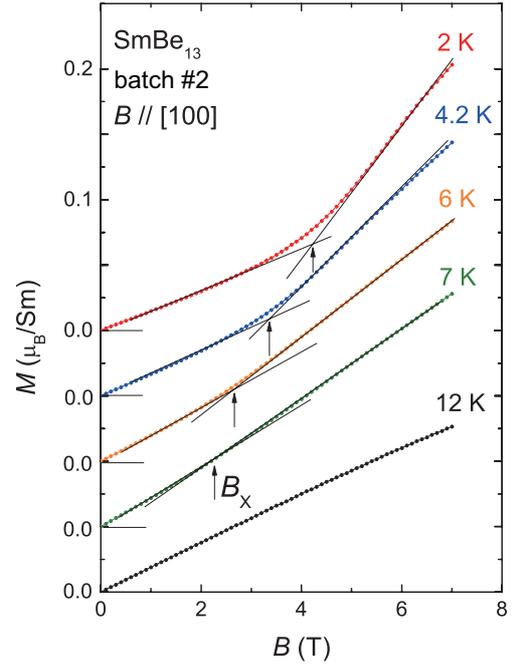}
\caption[]{\protect (Color online) Magnetization of SmBe$_{13}$ up to 7 T ($B$ // [100]) at several temperatures ($T$ = 2, 4.2, 6, 7, and 12 K). All the solid lines are guides to the eye, and the arrows indicate the intersection of two linear extrapolations, giving the definition of $B_{\rm X}$. 
}
\end{center}
\end{figure}

\subsection{Magnetic phase diagram}

The magnetic field--temperature ($B$--$T$) phase diagram of SmBe$_{13}$ for $B$ // [100], constructed from the $C$ (closed symbols) and $M$ (open symbols) measurements, is shown in Fig. 7. 
As the magnetic field increases to 9 T, $T_{\rm M}$ decreases from 8.3 to 7.7 K. 
The most striking feature is that the $B$--$T$ phase diagram of SmBe$_{13}$ consists of three regions below $T_{\rm M}$, indicating that the magnetic structure of this compound changes with the applied magnetic field. 
Here, we refer to these three regions as I, II, and III (see Fig. 7). 
Figure 7 also shows a contour plot of the difference in $M$($T$) between the ZFC and FC processes ($\equiv$ $\Delta$$M$), where the red color represents the largest $\Delta$$M$. 
This contour plot revealed that $\Delta$$M$ is present only in region III. 
Note that this magnetic phase diagram, including the difference between the ZFC and FC processes below $T_{\rm X}$, is consistent with that obtained from ultrasonic measurements under magnetic fields \cite{Mombetsu}.

\begin{figure}[htb]
\begin{center}
\includegraphics[width=0.9\linewidth]{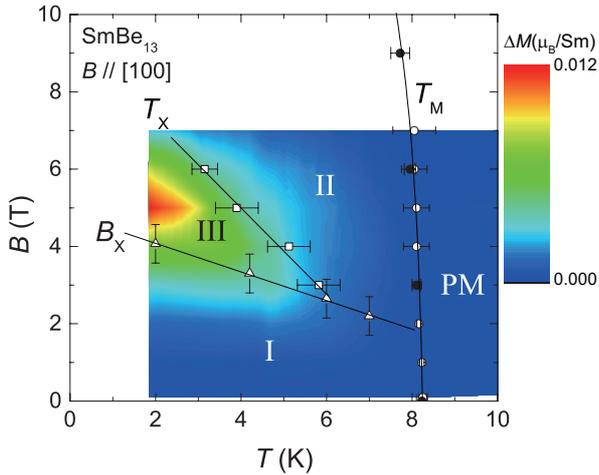}
\caption[]{\protect (Color online) $B$--$T$ phase diagram of SmBe$_{13}$ for $B$ // [100] axis. 
There are three regions (named I, II, and III) below $T_{\rm M}$. 
The solid lines are guides to the eye. 
In this figure, a contour plot of $\Delta$$M$ is also shown simultaneously. 
}
\end{center}
\end{figure}

\section{Discussion}

We now discuss a possible ordering state of SmBe$_{13}$.
Recently, novel orderings, which show an unusual magnetic-field response, have been found in some cage-structural Sm-based compounds; for example, the ordering temperature for SmRu$_4$P$_{12}$ increases with increasing magnetic field \cite{Matsuhira}, while SmTr$_2$Al$_{20}$ (Tr = Ti and Ta) show a rather magnetic-field-insensitive phase transition \cite{Higashinaka, Yamada}.
These behaviors are considered to be related to magnetic-octupole degrees of freedom in the $\Gamma_8$ quartet ground state \cite{Yoshizawa, Higashinaka, Yamada}. 
SmBe$_{13}$ also has the \mbox{cagelike} structure and the $\Gamma_8$ ground state; nevertheless such unusual field dependence of $T_{\rm M}$ cannot be observed \cite{Mombetsu}. 
In addition, our recent muon-spin relaxation measurements of SmBe$_{13}$ indicate that an internal magnetic field larger than 0.1 T occurs at a muon stopping site below $T_{\rm M}$ \cite{Tabata}, suggesting that the primary order parameter for SmBe$_{13}$ is a magnetic dipole. 
Thus, the simplest explanation for the ordering state of SmBe$_{13}$ is an AFM ordering, since the $\chi$($T$) curve shows the cusp anomaly at $T_{\rm M}$, and $T_{\rm M}$ decreases by applying a magnetic field. 
However, the occurrence of the simple AFM ordering appears to contradict the presence of the dominant FM interaction in the PM state, suggesting that the ordered state of SmBe$_{13}$ has a rather complex magnetic structure. 
High-field ultrasonic measurements up to 61.3 T revealed that the ordered state of SmBe$_{13}$ is suppressed by a magnetic field of 43 T for $B$ // [100] axis. \cite{Mombetsu} 
The obtained critical magnetic field deviates from the value estimated from the mean-field calculation assuming the $\Gamma_8$ CEF ground state and a simple G-type AFM state. \cite{Mombetsu}

We suggest that the most plausible candidate for the ordered state of SmBe$_{13}$ is a helical-magnetic ordering. 
It is noteworthy that many MBe$_{13}$ compounds with M = Gd -- Er and Np order into a helical-magnetic structure with propagation vector \mbox{\boldmath $Q$} $\sim$ [00$\frac{1}{3}$] at the transition temperature $T_{\rm heli}$ \cite{Bouree, Hiess}. 
They also show both an AFM-like cusp anomaly at $T_{\rm heli}$ in $\chi$($T$) and a positive $\theta_{\rm CW}$, whose absolute value is comparable to $T_{\rm heli}$. 
These features are similar to those in SmBe$_{13}$. 
In the heavy-rare-earth MBe$_{13}$ compounds, the occurrence of the helical ordering is explained by competition between exchange interactions between the first- and second-neighboring (100) planes, where each plane has a strong FM interaction \cite{Becker}. 
Here, we defined the $c$-axis as the helical axis of the magnetic structure in a cubic crystal for convenience. 
In this model, the exchange interactions between the first- and second-neighboring $c$ planes were assumed to be FM and AFM, respectively. 
Since these MBe$_{13}$ compounds are metallic systems with well-localized $f$ electrons \cite{Bucher, Dervenagas}, their magnetic interactions are considered to originate from the Ruderman--Kittel--Kasuya--Yosida (RKKY) interaction. 
In this context, the competition between the exchange interactions may be due to the oscillation of spin polarization, namely, the RKKY oscillation.

In addition, the magnetic phase diagram of SmBe$_{13}$ constructed from this study is similar to that reported previously for the helical magnet HoBe$_{13}$ \cite{Dervenagas}. 
In this system, neutron diffraction measurements have revealed that two magnetic transitions are successively induced below $T_{\rm heli}$ by applying a magnetic field along the $c$-axis; the first one is from the helical structure with three domains to a single-domain conical structure, and the second one is from the conical structure to a two-domain canted magnetic structure \cite{Dervenagas}. 
If SmBe$_{13}$ has similar magnetic structures below $T_{\rm M}$, the multiple regions in the phase diagram can be understood. 
The change in slope of the $M$($B$) curve and the finite $\Delta$$M$ in region III of SmBe$_{13}$ could be explained by the presence of the spontaneous FM component derived from the conical structure with magnetic domains. 
Such a difference in $M$($T$) between the FC and ZFC processes has also been observed in a conical-ordered phase of Pd$_3$Mn. \cite{Pd3Mn} 
However, since these suggestions are based solely on the results of $M$ measurements, we should not exclude the possibility of other magnetic structures accompanying the spontaneous FM components, for instance, a fan structure, canted AFM structure, and uncompensated AFM structure.  
Furthermore, the $M$($B$) curve of SmBe$_{13}$ does not show a metamagnetic transition near the region boundary, which is different from the cases of HoBe$_{13}$ and the typical helical magnet MnP. \cite{MnP} 
To clarify the magnetic structures below $T_{\rm M}$ of SmBe$_{13}$, we need further detailed studies, particularly microscopic measurements, such as neutron scattering on isotope-substituted samples and nuclear magnetic resonance.

If SmBe$_{13}$ also exhibits the helical-magnetic ordering at $T_{\rm M}$, it will be the first collateral evidence that the helical-ordering state is a common ground state in the magnetic MBe$_{13}$ compounds including the light-rare-earth systems. 
Note that the ground state of UBe$_{13}$ is not the helical ordering but the unconventional SC. 
In addition, the inelastic neutron scattering measurements for UBe$_{13}$ have revealed the development of AFM short-range magnetic correlations with the propagation vector \mbox{\boldmath $q$} = [$\frac{1}{2}$$\frac{1}{2}$0] at low temperatures \cite{Coad}. 
It is necessary to elucidate the reason why only UBe$_{13}$ possesses a different ground state and magnetic correlation from those in other magnetic MBe$_{13}$ compounds.

\section{Summary} 
We have succeeded in growing single crystals of SmBe$_{13}$, and we reported the results of $C$, $M$, and $\rho$ measurements using the grown single crystals. 
From this study, we obtained the following results concerning a PM state: (i) 4$f$ electrons of Sm$^{3+}$ ions are well localized, (ii) a plausible CEF level scheme is $\Gamma_8$ -- $\Gamma_7$ ($\Delta_{\rm CEF}$ = 90 K), and (iii) the dominant magnetic interaction between the 4$f$-dipole moments is FM on the whole. 
Interestingly, $\chi$($T$) shows an AFM-like cusp anomaly at $T_{\rm M}$ $\sim$ 8.3 K despite the presence of the dominant FM interaction ($\theta_{\rm CW}$ $\sim$ 10.8 K). 
Similar features have also been found in other isostructural MBe$_{13}$ compounds showing helical-magnetic ordering. 
Furthermore, a magnetic phase diagram with multiple regions of SmBe$_{13}$ was constructed from the $C$ and $M$ measurements under a magnetic field. 
The obtained phase diagram seems to be similar to that of the isostructural system HoBe$_{13}$, which shows the helical ordering. 
These findings suggest that SmBe$_{13}$ undergoes a helical or similar magnetic ordering at $T_{\rm M}$.

\begin{acknowledgment}

The present research was supported by JSPS KAKENHI Grant Numbers JP20224015(S), JP25400346(C), JP26400342(C), JP15H05882, and JP15H05885(J-Physics).

\end{acknowledgment}

\end{document}